\documentclass[twocolumn,preprintnumbers,showpacs,pre,floatfix,amsmath,amssymb,notitlepage]{revtex4-1}
\usepackage{epsfig}
\usepackage{graphicx}
\usepackage{amsmath}
\usepackage{color}
\usepackage{amssymb}
\usepackage{setspace}
\usepackage{dcolumn}
\usepackage{bm}
\usepackage{times}
\usepackage{enumerate}
\usepackage{footnote}
\input epsf

\graphicspath{{figures/}}

\begin{document}

\author{Per Sebastian Skardal}
\email{skardals@gmail.com} 
\affiliation{Departament d'Enginyeria Inform\`{a}tica i Matem\`{a}tiques, Universitat Rovira i Virgili, 43007 Tarragona, Spain}

\author{Alex Arenas}
\email{alexandre.arenas@urv.cat}
\affiliation{Departament d'Enginyeria Inform\`{a}tica i Matem\`{a}tiques, Universitat Rovira i Virgili, 43007 Tarragona, Spain}

\title{Disorder induces explosive synchronization}

\begin{abstract}
We study explosive synchronization, a phenomenon characterized by first-order phase transitions between incoherent and synchronized states in networks of coupled oscillators. While explosive synchronization has been the subject of many recent studies, in each case strong conditions on either the heterogeneity of the network, its link weights, or its initial construction are imposed to engineer a first-order phase transition. This raises the question of how robust explosive synchronization is in view of more realistic structural and dynamical properties. Here we show that explosive synchronization can be induced in mildly heterogeneous networks by the addition of quenched disorder to the oscillators' frequencies, demonstrating that it is not only robust to, but moreover promoted by, this natural mechanism. We support these findings with numerical and analytical results, presenting simulations of a real neural network as well as a self-consistency theory used to study synthetic networks.
\end{abstract}


\maketitle

\section{Introduction}\label{sec1}

Phase transitions and critical phenomena are central topics in the research of complex networks because of their deep implications in dynamical processes~\cite{Dorogovtsev2008RMP}. Recently, explosive, i.e., very sharp, phase transitions have garnered a great deal of attention from scientists, first arising in the context of network percolation~\cite{Achlioptas2009Science}. While these transitions were eventually proven to be continuous, and thus not explosive~\cite{daCosta2010PRL,Riordan2011Science}, interest in abrupt phase transitions was reignited in the context of synchronization~\cite{Arenas2008PR}. Synchronization has long served as a major tool in studying emergent collective behavior in ensembles of coupled dynamical agents~\cite{Strogatz2003,Pikovsky2003}, with examples found in nature, e.g., rhythmic flashing of fireflies~\cite{Buck1988QRB} and mammalian circadian rhythms~\cite{Glass1988}, in engineering, e.g., power grids~\cite{Motter2013NaturePhys} and oscillations of pedestrian bridges~\cite{Strogatz2005Nature}, and at their intersection, e.g., synthetic cell engineering~\cite{Prindle2012Nature}. In particular, the Kuramoto model has served as a paradigm for both modeling and understanding synchronization~\cite{Kuramoto1984}. When placed on a network, the Kuramoto model consists of an ensemble of $N$ phase oscillators, $\theta_i$ for $i=1,\dots,N$, whose evolution is governed by
\begin{align}
\dot{\theta}_i = \omega_i + \lambda\sum_{j=1}^NA_{ij}\sin\left(\theta_j-\theta_i\right),\label{eq:Kuramoto}
\end{align}
where $\omega_i$ is the natural (intrinsic) frequency of oscillator $i$, $\lambda$ is the global coupling strength, and the adjacency matrix $[A_{ij}]$ encodes the network topology that defines the oscillators' interactions. 

In 2011, G\'{o}mez-Garde\~{n}es et al.~\cite{GomezGardenes2011PRL} found that for sufficiently heterogeneous network topologies (e.g., those generated with the Barab\'{a}si-Albert preferential attachment model~\cite{Barabasi1999Science}) a simple degree-frequency correlation defined by $\omega_i=k_i$, where $k_i=\sum_{j}A_{ij}$ is the degree of node $i$, induces an explosive phase transition in the order parameter $r$, defined by
\begin{align}
re^{i\psi}=\frac{1}{N}\sum_{j=1}^Ne^{i\theta_j},\label{eq:Order}
\end{align}
as the coupling strength $\lambda$ is varied. In particular, $re^{i\psi}$ represents the centroid of all oscillators when placed on the complex unit circle, with $r\approx0$ and $r\approx1$ indicating incoherent and synchronized behavior, respectively. Such explosive synchronization is characterized by the emergence of a range of coupling strengths where incoherent and synchronized states are both stable, unlike the first-order transition studied in Ref.~\cite{Pazo2005PRE}. Subsequently, significant attention has been paid to the further exploration of degree-frequency correlations~\cite{Skardal2013EPL,Liu2013EPL,Sonnenschein2013EPJB,Coutinho2013PRE} and in particular explosive synchronization~\cite{Leyva2012PRL,Peron2012PRE1,Peron2012PRE2,Chen2013Chaos,Ji2013PRL,Leyva2013SR,Leyva2013PRE,Li2013PRE,Su2013EPL,Zhang2013PRE,Zhu2013PRE,Zou2014PRL}. While this research has augmented our understanding of explosive synchronization and its relationship with dynamical and structural correlations, in each case strong conditions are necessarily imposed on either the heterogeneity of the network, its link weights, or its initial construction to engineer first-order phase transitions. This raises the following question: How robust is the phenomenon of explosive synchronization in the view of more realistic dynamical and structural properties? In this paper we demonstrate that explosive synchronization can be induced in both real and synthetic networks by the addition of quenched disorder to the oscillators' frequencies. In particular, the resulting networks consist of oscillators whose frequencies are correlated with, but not precisely determined by, local structural properties; a property one might expect to find in many real-world situations. Therefore, we conclude that, explosive synchronization is not only robust to, but moreover promoted by, this very natural mechanism. 

\begin{figure*}[t]
\centering
\epsfig{file =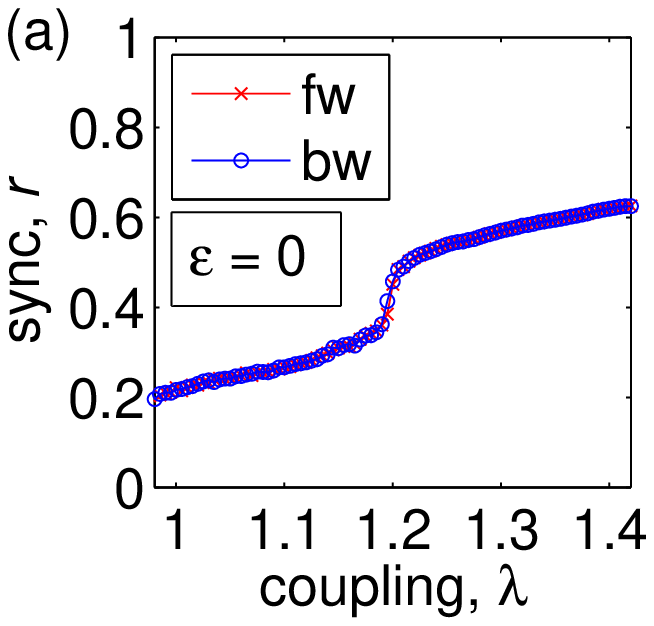, clip =,width=0.25\linewidth }
\epsfig{file =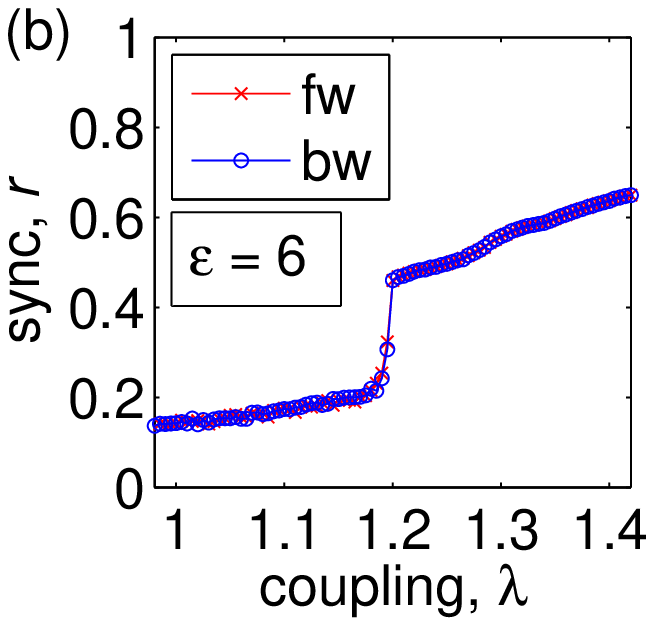, clip =,width=0.25\linewidth }
\epsfig{file =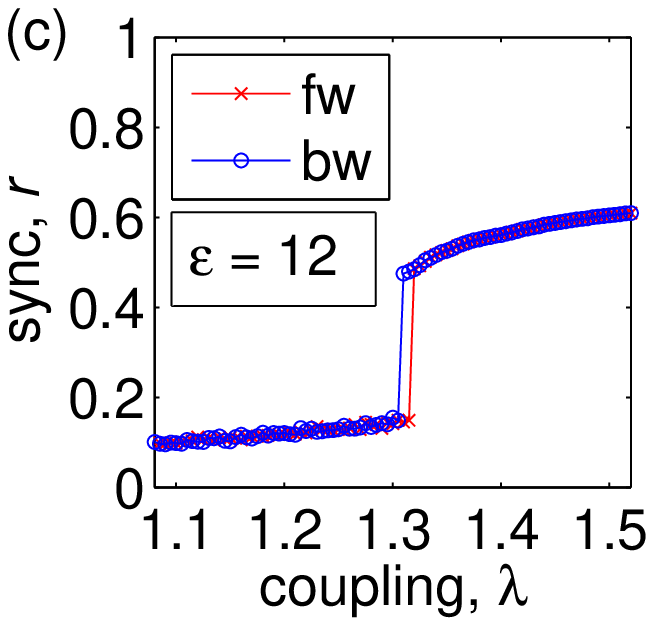, clip =,width=0.25\linewidth }
\caption{(Color online) C. elegans neural network dynamics: (a)--(c) Forward (red crosses) and backward (blue circles) synchronization profiles $r$ vs $\lambda$ for the C. elegans neural network with increasing disorder: $\epsilon=0$, $6$, and $12$. See Appendix~\ref{appA} for details of the network.} 
\label{fig:CE}
\end{figure*}

The remainder of this paper is organized as follows. In Sec.~\ref{sec2} we describe the quenched disorder considered in this paper and present the resulting dynamics on a real network. In particular, we consider the dynamics of Eq.~(\ref{eq:Kuramoto}) on the C. elegans neural network. In Sec.~\ref{sec3} we present the derivation of a self-consistency condition that describes the steady-state dynamics of large networks. In Sec.~\ref{sec4} we present numerical and analytical results, comparing results from direct simulations of Eq.~(\ref{eq:Kuramoto}) to the solutions of the derived self-consistency condition. Finally, in Sec.~\ref{sec5} we conclude with a discussion of our results.

\section{Quenched Disorder and Dynamics of a Real Network}\label{sec2}

In this paper we consider the dynamics of Eq.~(\ref{eq:Kuramoto}) and the degree of synchronization $r$ defined by Eq.~(\ref{eq:Order}) for a particular choice of frequencies. Specifically, we set each natural frequency equal to its corresponding nodal degree plus some randomness, i.e., $\omega_i=k_i+\xi_i$. We assume that the randomness $\xi_i$ for each oscillator $i$ is independent from all other oscillators and for simplicity assume that it is drawn from the uniform distribution $\mathcal{U}(-\epsilon,\epsilon)$. Thus, the parameter $\epsilon\ge0$ controls the amount of disorder added to the frequencies. We note that in the limit as $\epsilon\to0^+$ we recover the simple case $\omega_i=k_i$ studied in Ref.~\cite{GomezGardenes2011PRL}.

To illustrate the effect of this choice of frequencies, we present results from simulating Eq.~(\ref{eq:Kuramoto}) on the C. elegans neural network. The C. elegans neural network is a real network composed of 302 nodes (neurons) with a mildly heterogeneous degree distribution that undergoes synchronization dynamics~\cite{Chen2006PNAS}. For the details of the dataset, see Appendix~\ref{appA}. For our purposes we interpret each link in the network to be undirected and unweighted, i.e., the adjacency matrix $A$ is symmetric and $A_{ij}=1$ if a link connects nodes $i$ and $j$, otherwise $A_{ij}=0$. 

In Figs.~\ref{fig:CE}(a)--(c) we present the results, plotting the forward (fw) and backward (bw) synchronization profiles $r$ vs $\lambda$ for $\epsilon=0$ (i.e., no disorder), $\epsilon=6$, and $\epsilon=12$. Each profile is obtained by first slowly increasing, then slowly decreasing $\lambda$. For each value of $\lambda$ we simulate a long transient to reach steady state then average $r$ over a significant amount of time. We note the following interesting behavior. First, in the absence of disorder ($\epsilon=0$) the transition from incoherent to synchronized dynamics is second-order, i.e., not explosive, and relatively mild. For $\epsilon=6$ the transition becomes much sharper, however remains second-order. Finally, for $\epsilon=12$ the transition to synchronization is explosive as indicated by the emergence of a thin region of bistability. We emphasize here that the underlying network is the same in each panel of Fig.~\ref{fig:CE}. Thus, for a network whose transition to synchronization is second-order in the absence of disorder, a sufficient amount of disorder induces explosive synchronization.

\section{Self-consistency Analysis}\label{sec3}

We now present the derivation of a self-consistency condition that we will use to obtain analytical results. Following Ref.~\cite{Skardal2013EPL} we consider the thermodynamic limit of large networks $N\to\infty$ whose degrees and frequencies can be described by a joint probability distribution $P(k,\omega)$. Additionally, we assume that no structural correlations exist in the network. Under these assumptions, we search for synchronized solutions consisting of a single synchronized cluster traveling with angular velocity $\Omega$. We note that $\Omega$ can be reasonably approximated by the mean natural frequency $\sum_i\omega_i/N$, however, the following self-consistency analysis defines $\Omega$ along with $r$. We begin by entering a rotating reference frame by introducing the change of variables $\phi_i=\theta_i-\Omega t$, which transforms Eq.~(\ref{eq:Kuramoto}) into
\begin{align}
\dot{\phi}_i = (\omega_i-\Omega)+\lambda\sum_{j=1}^NA_{ij}\sin(\phi_j-\phi_i).\label{eq:Theory01}
\end{align}
We next introduce the set of local order parameters defined by
\begin{align}
r_ie^{i\psi_i}=\sum_{j=1}^NA_{ij}e^{i\phi_j},\label{eq:Theory02}
\end{align}
whose magnitude $r_i$ can be interpreted as a measure of synchronization among the network neighbors of node $i$. We note that precisely $k_i$ terms contribute to $r_i$, and thus $r_i\in[0,k_i]$, as opposed to the global order parameter $r\in[0,1]$ as defined in Eq.~(\ref{eq:Order}). Importantly, Eq.~(\ref{eq:Theory02}) simplifies Eq.~(\ref{eq:Theory01}) to
\begin{align}
\dot{\phi}_i=(\omega_i-\Omega)+\lambda r_i\sin(\psi_i-\phi_i),\label{eq:Theory03}
\end{align}
which can be used to classify the dynamics of each oscillator given $r_i$, $\psi_i$, $\omega_i$, $\lambda$, and $\Omega$. If $|\omega_i-\Omega|\le\lambda r_i$, then $\phi_i$ reaches a fixed point defined by $\sin(\phi_i-\psi_i)=(\omega_i-\Omega)/\lambda r_i$, indicating that it becomes phase-locked. Otherwise, $\phi_i$ never reaches a fixed point, indicating that oscillator $i$ drifts for all time.

To classify the degree of synchronization we now inspect the local order parameters. In principle, the contribution to each local order parameter can be divided into that from the phase-locked and drifting oscillators, i.e., $r_i = r_i^{lock}+r_i^{drift}$, however for simplicity we neglect the drifting contribution and consider only the contribution of locked oscillators satisfying $|\omega_i-\Omega|\le\lambda r_i$. In Appendix~\ref{appB} we present a derivation for the contribution of the drifting oscillators, but note here that we find this contrition is second-order in comparison to that of the locked oscillators. Thus, we approximate each local order parameters as
\begin{align}
r_i=\sum_{|\omega_j-\Omega|\le\lambda rj}A_{ij}e^{i(\phi_j-\psi_i)},\label{eq:Theory04}
\end{align}
noting that we sum only over neighbors that satisfy the phase-locking criteria. 

Next we make two additional simplifying approximations regarding the local order parameters. First, we assume that all the local average phases are approximately equal, i.e., $\psi_i\approx\psi_j$ for all $(i,j)$ pairs. This is a reasonable assumption given that a single synchronized cluster exists, as is typically the case for networks without strong modular structure. Second, recalling that exactly $k_i$ terms contribute to $r_i$, we propose that each $r_i$ is approximately proportional to $k_i$, i.e., there exists some $\tilde{r}$ such that $r_i\approx \tilde{r}k_i$ for all $i$. Both assumptions have been used and numerically validate in previous synchronization studies~\cite{Skardal2013EPL,Ichinomiya2004PRE,Restrepo2005PRE} and are predicted to be most accurate for networks with relatively large mean degree, such that fluctuations in the local averages sufficiently diminish. It can also be shown that the constant $\tilde{r}$ is a good approximation for the global degree of synchronization $r$, and thus we approximate $r\approx\tilde{r}$. In Appendix~\ref{appC} we present some numerical experiments validating these important approximations.

Taking into account these approximations, we expand the exponential in Eq.~(\ref{eq:Theory04}) into cosine and sine and use that at steady-state phase-locked oscillators satisfy $\sin(\phi_i-\psi_i)=(\omega_i-\Omega)/\lambda r_i$ to obtain
\begin{align}
rk_i=\sum_{|\omega_j-\Omega|\le\lambda rk_j}A_{ij}\left[\sqrt{1-\left(\frac{\omega_j-\Omega}{\lambda rk_j}\right)^2}+i\frac{\omega_j-\Omega}{\lambda rk_j}\right].\label{eq:Theory05}
\end{align}
Summing Eq.~(\ref{eq:Theory05}) over $i$, dividing by $N$, and separating into real and imaginary parts, we finally arrive, after some rearranging, at
\begin{align}
r &= \frac{\langle k\rangle^{-1}}{N}\sum_{|\omega_j-\Omega|\le\lambda rk_j}k_j\sqrt{1-\left(\frac{\omega_j-\Omega}{\lambda rk_j}\right)^2},\label{eq:Theory06}\\
\Omega &= \frac{\sum_{|\omega_j-\Omega|\le\lambda rk_j}\omega_j}{\sum_{|\omega_j-\Omega|\le\lambda rk_j}1}.\label{eq:Theory07}
\end{align}
In principle, given a sequence of degree-frequency pairs $\{(k_i,\omega_i)\}_{i=1}^N$, Eqs.~(\ref{eq:Theory06}) and (\ref{eq:Theory07}) can be solved self-consistently for $r$ and $\Omega$. However, to obtain results for entire ensembles of networks described by a general join distribution $P(k,\omega)$, we note that in the large $N$ limit Eqs.~(\ref{eq:Theory06}) and (\ref{eq:Theory07}) can be transformed into the integrals
\begin{align}
r &= \langle k\rangle^{-1}\iint_{|\omega-\Omega|\le\lambda rk} P(k,\omega)k\sqrt{1-\left(\frac{\omega-\Omega}{\lambda rk}\right)^2}\mathrm{d}\omega\mathrm{d}k,\label{eq:Self01}\\
\Omega &=\frac{\iint_{|\omega-\Omega|\le\lambda rk} P(k,\omega)\omega\mathrm{d}\omega\mathrm{d}k}{\iint_{|\omega-\Omega|\le\lambda rk} P(k,\omega)\mathrm{d}\omega\mathrm{d}k}.\label{eq:Self02}
\end{align}

Before proceeding, we make a few important remarks. First, Eqs.~(\ref{eq:Self01}) and (\ref{eq:Self02}) give a self-consistency condition for the steady-state order parameter $r$ and the angular velocity of the synchronized population $\Omega$. In principle, only the joint distribution $P(k,\omega)$ is needed to solve the self-consistency condition, so a given solution is valid for the entire family of networks represented by $P(k,\omega)$. Second, we find that the solutions of Eqs.~(\ref{eq:Self01}) and (\ref{eq:Self02}) match up well with the solutions of the analogous sums in Eqs.~(\ref{eq:Theory06}) and (\ref{eq:Theory07}) for finite sequences of degree-frequency pairs $\{(k_i,\omega_i)\}_{i=1}^N$. We illustrate this with some numerical investigations in Appendix~\ref{appC}. Finally, in cases where frequencies are defined precisely by their degree, i.e., $\omega_i=\omega(k_i)$, the joint distribution $P(k,\omega)$ is defined in terms of delta functions and the double integrals in Eqs.~(\ref{eq:Self01}) and (\ref{eq:Self02}) reduce to single integrals.

\section{Results}\label{sec4}

We next present numerical and analytical results for two classes of synthetic networks, comparing simulations of Eq.~(\ref{eq:Kuramoto}) to solutions of the self-consistency condition given by Eqs.~(\ref{eq:Self01}) and (\ref{eq:Self02}). Additionally, we utilize the self-consistency condition to calculate the phase diagram for both classes of networks. In particular, we consider (i) stretched exponential (SE) and (ii) scale-free (SF) networks whose degree distributions are given by $P(k)\propto k^{\beta-1}\exp[-(k/\mu)^\beta]$ and $P(k)\propto k^{-\gamma}$, respectively. In both cases we choose parameters such that the degree distributions are mildly heterogeneous, falling between very heterogeneous SF networks with $\gamma<3$~\cite{Clauset2009SIAM} and very homogeneous Erd\H{o}s-R\'{e}nyi networks~\cite{Erdos1960}.

\subsection{Stretched exponential networks}

\begin{figure}[t]
\centering
\epsfig{file =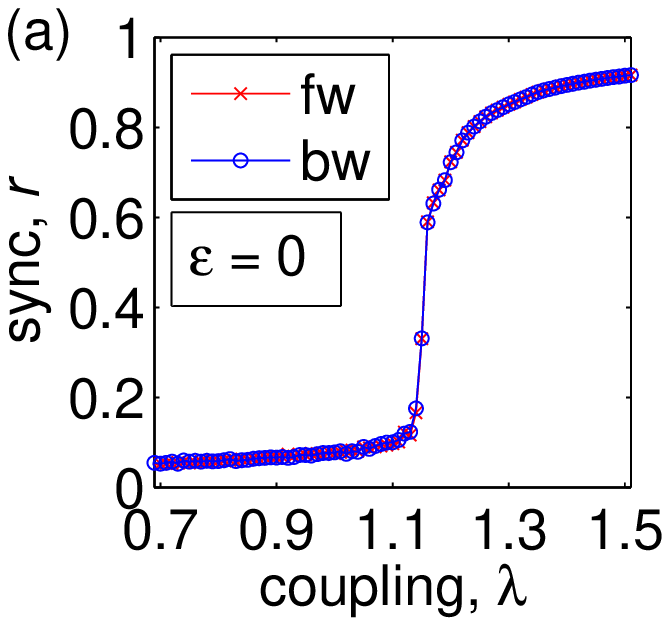, clip =,width=0.48\linewidth }
\epsfig{file =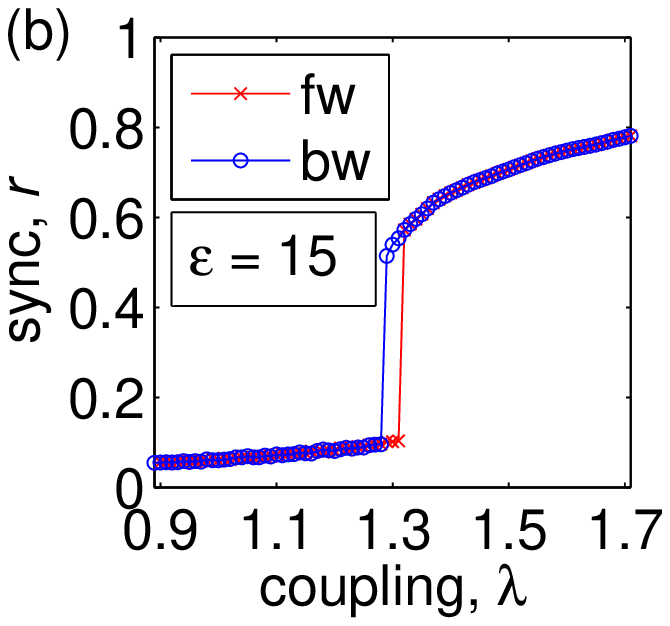, clip =,width=0.48\linewidth } \\
\epsfig{file =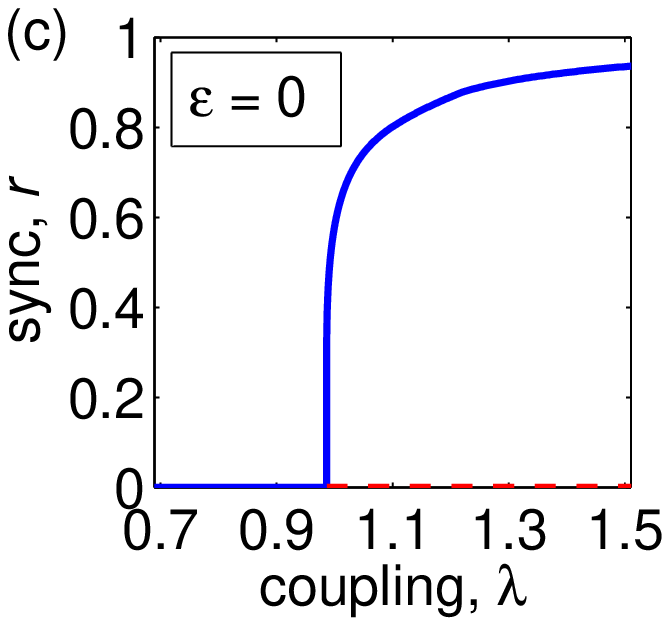, clip =,width=0.48\linewidth }
\epsfig{file =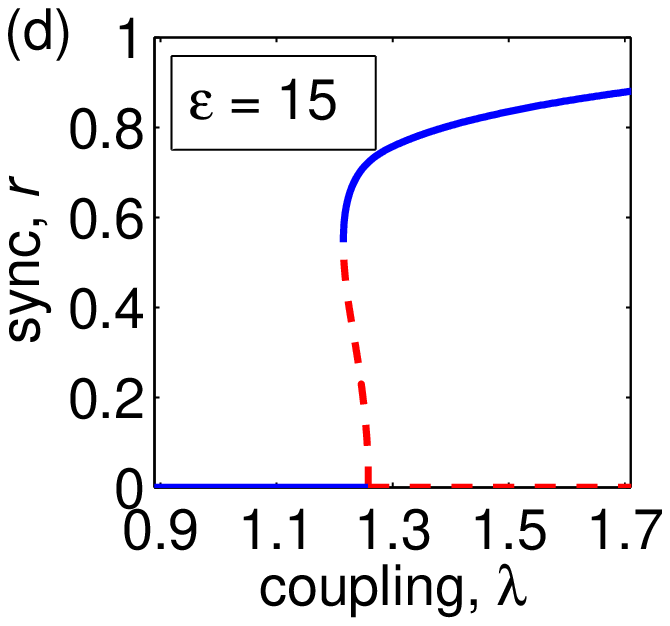, clip =,width=0.48\linewidth }
\caption{(Color online) SE networks: (a)--(b) Forward (fw) and backward (bw) synchronization profiles $r$ vs $\lambda$ from direct simulation of a SE network of size $N=1000$ for $\epsilon=0$ and $15$. (c)--(d) Synchronization profiles $r$ vs $\lambda$ given by solution curves of Eqs.~(\ref{eq:Self01}) and (\ref{eq:Self02}) for $\epsilon=0$ and $15$. Other parameters are $\mu=5$, $\beta=0.7$, and $k_0=10$.} \label{fig:HystSE}
\end{figure}

First we focus on SE networks, characterized by degree distribution $P(k)\propto k^{\beta-1}\exp[-(k/\mu)^\beta]$. We note that the choice $\beta=1$ yields the typical exponential distribution, while $\beta<1$ and $\beta>1$ stretches and compresses the distribution, respectively. Here we choose $\beta=0.7$ and $\mu=5$, and impose a minimum degree of $k_0=10$. In Figs.~\ref{fig:HystSE}(a) and \ref{fig:HystSE}(b) we plot the forward and backward synchronization profiles from direct simulation of Eq.~(\ref{eq:Kuramoto}) for $\epsilon=0$ and $15$ on a network of size $N=1000$ constructed using the configuration model~\cite{Bekessy1972}. We note that, similar to the C. elegans network, in the absence of disorder the transition from incoherence to synchronization is second-order, but with enough disorder the transition becomes explosive with a clear bistable regime emerging.

To complement these simulations, in Fig.~\ref{fig:HystSE}(c) and \ref{fig:HystSE}(d) we plot solution curves of the self-consistency condition in Eqs.~(\ref{eq:Self01}) and (\ref{eq:Self02}) for the same $\epsilon$ values, calculated numerically. We deduce the stability of each branch using the topology of the curves, plotting stable and unstable branches in solid blue and dashed red, respectively. Results are plotted with the same horizontal axis range and stacked vertically with corresponding $\epsilon$ values of the simulation results in Figs.~\ref{fig:HystSE}(a) and \ref{fig:HystSE}(b) for easy comparison. We note first that the self-consistency condition correctly predicts a second-order transition for no disorder ($\epsilon=0$) and a first-order transition for $\epsilon=15$ via the formation of a bistable region. Additionally, while the critical coupling values indicating transition points predicted by the self-consistency condition come slightly earlier than those observed in the simulations, they provide a reasonable prediction nonetheless. This offset is most likely due to the average phases $\psi_i$ being approximately equal rather than precisely equal, as assumed in the derivation of Eqs.~(\ref{eq:Self01}) and (\ref{eq:Self02}).

\begin{figure}[t]
\centering
\epsfig{file =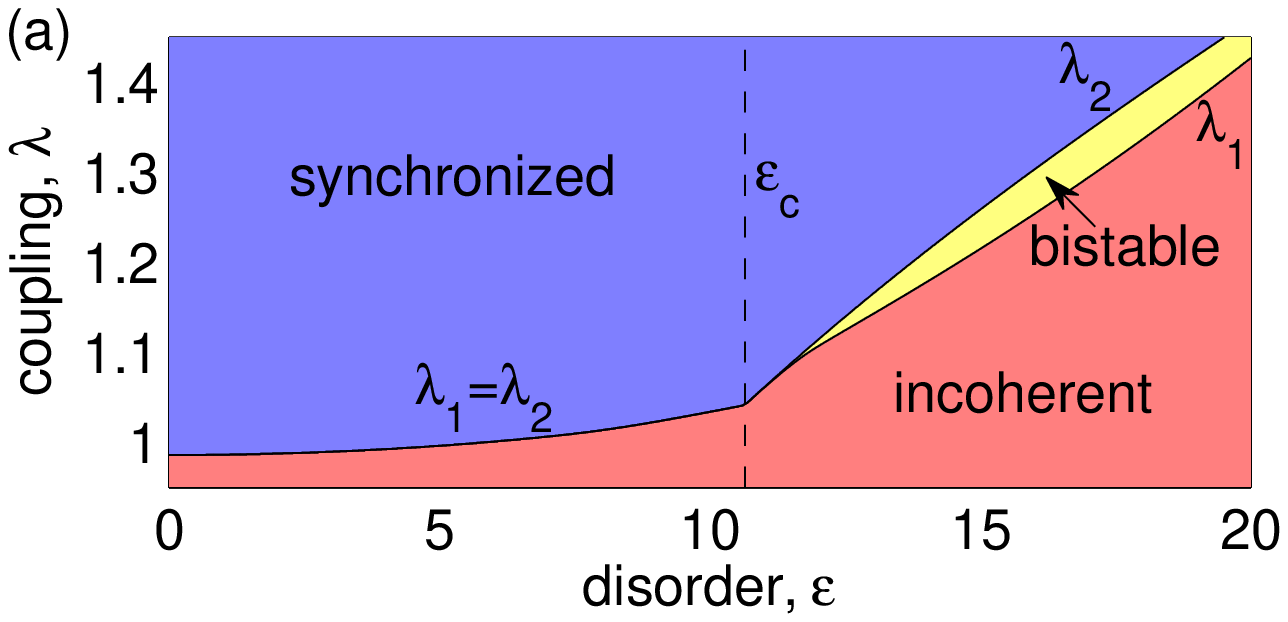, clip =,width=0.96\linewidth }
\epsfig{file =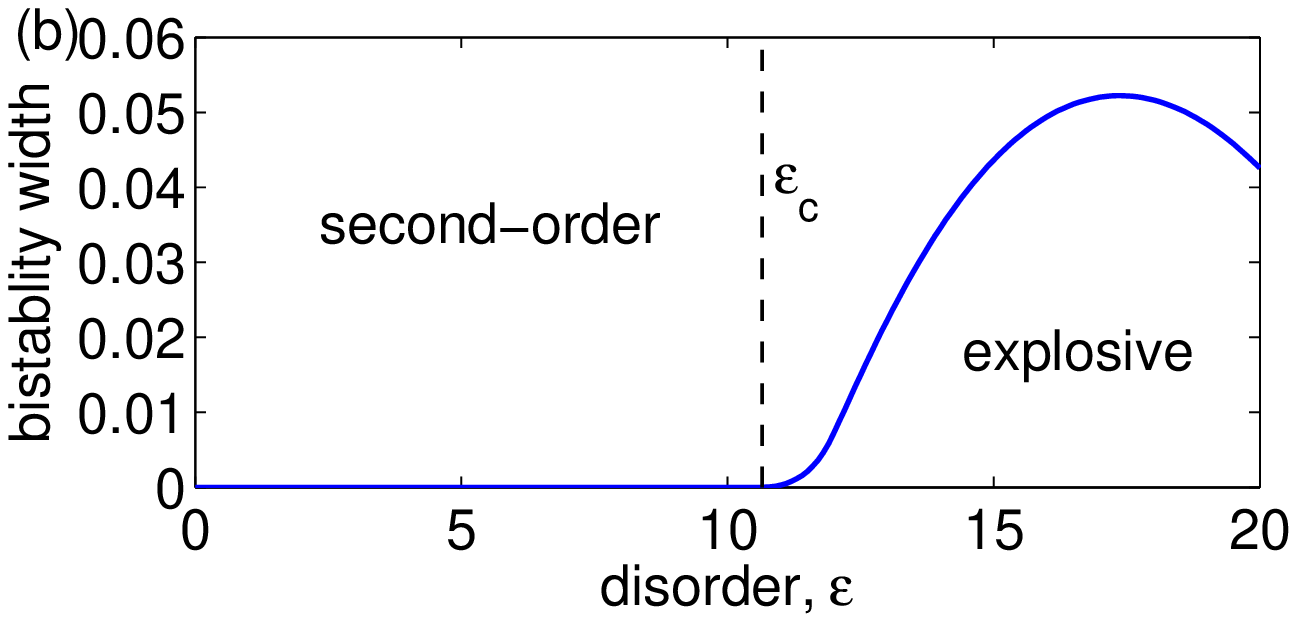, clip =,width=0.96\linewidth }
\caption{(Color online) Phase space of SE networks: (a) Phase space with critical coupling strengths $\lambda_1$ and $\lambda_2$ separating incoherent, synchronized, and bistable regions and (b) the bistability width $\lambda_2-\lambda_1$ versus $\epsilon$ computed from Eqs.~(\ref{eq:Self01}) and (\ref{eq:Self02}). Other parameters are $\mu=5$, $\beta=0.7$, and $k_0=10$.} \label{fig:PhaseSpaceSE}
\end{figure}

Equations~(\ref{eq:Self01}) and (\ref{eq:Self02}) also allow us to compute the phase space for networks with a prescribed degree-frequency distribution by calculating, as a function of the disorder intensity $\epsilon$, the critical points $\lambda_1$ and $\lambda_2$ corresponding to the birth of the synchronized branch and the collision point of the synchronized and incoherent branches. In particular, second-order phase transitions are characterized by $\lambda_1=\lambda_2$ separating the incoherent and synchronized regions, while explosive phase transitions are characterized by $\lambda_1<\lambda_2$, yielding a bistable region in between the incoherent and synchronized regions. In Fig.~\ref{fig:PhaseSpaceSE}(a) we plot the phase space for SE networks with $\beta=0.7$, $\mu=5$, and $k_0=10$. In Fig.~\ref{fig:PhaseSpaceSE}(b) we also plot the bistability width $\lambda_2-\lambda_1$ as a function of $\epsilon$. Here we see that for small disorder values $\epsilon$ the transition from incoherent to synchronized regions is second-order with $\lambda_1=\lambda_2$ until at $\epsilon_c\approx10.65$ (denoted with a vertical dashed line), a bistable region is born with $\lambda_1<\lambda_2$.

\subsection{Scale-free networks}

Next we consider SF networks, characterized by the degree distribution $P(k)\propto k^{-\gamma}$. In Ref.~\cite{GomezGardenes2011PRL} G\'{o}mez-Garde\~{n}es et al. showed that for sufficiently small $\gamma$ (e.g., $\gamma\le3$) the correlation $\omega_i=k_i$ (i.e., no disorder added to the frequencies) is sufficient to induce explosive synchronization. However, for less heterogeneous networks, i.e., $\gamma>3$, explosive synchronization is lost for the assignment $\omega_i=k_i$. Thus, we consider here the latter case of more mildly heterogeneous SF networks with $\gamma>3$ with quenched disorder added to the frequencies. Specifically, we choose $\gamma=3.5$ and impose a minimum degree of $k_0=10$.

\begin{figure}[t]
\centering
\epsfig{file =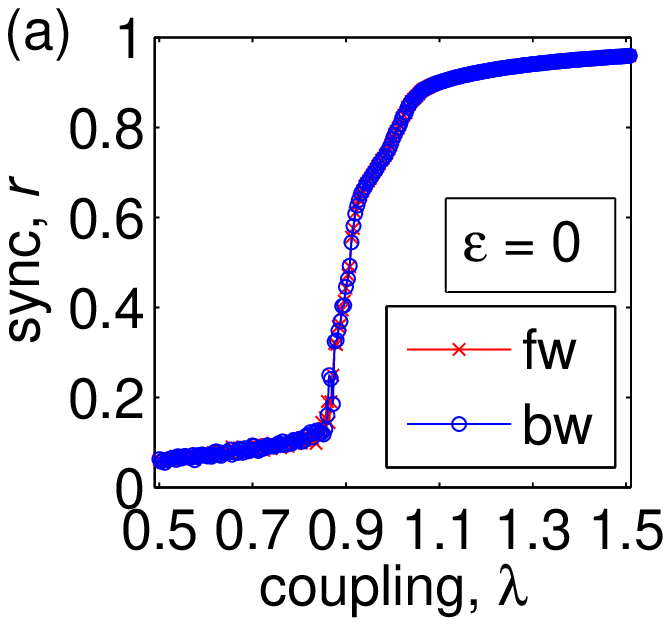, clip =,width=0.48\linewidth }
\epsfig{file =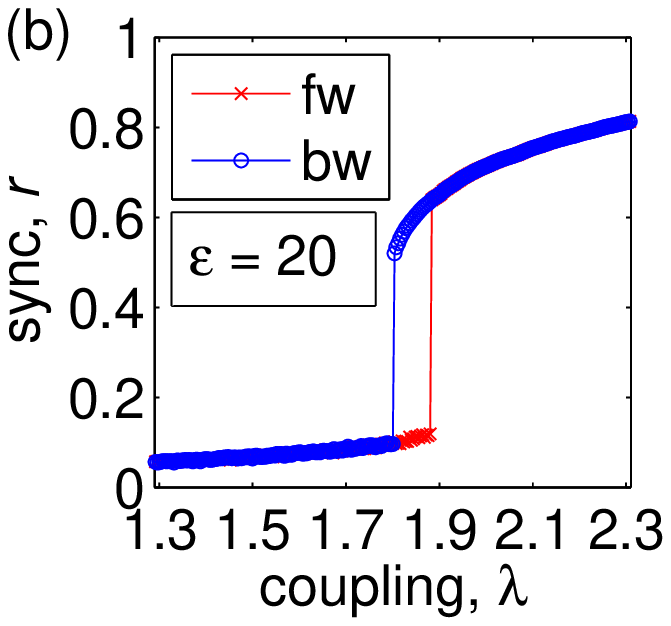, clip =,width=0.48\linewidth } \\
\epsfig{file =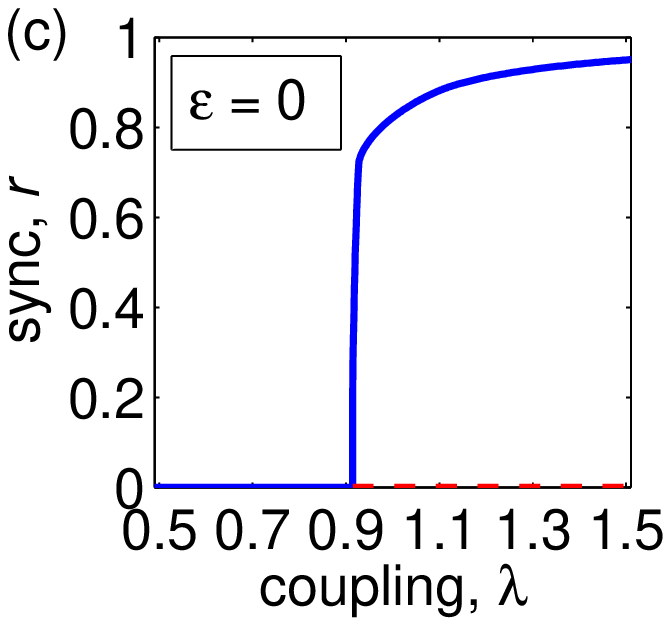, clip =,width=0.48\linewidth }
\epsfig{file =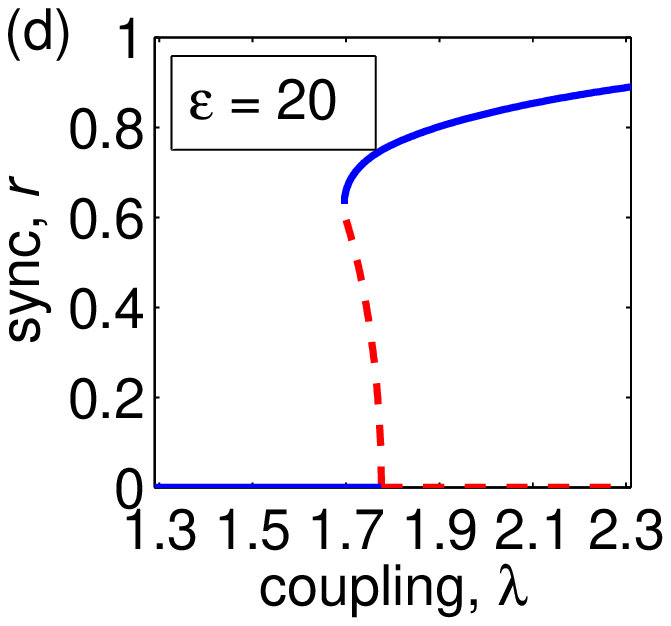, clip =,width=0.48\linewidth }
\caption{(Color online) SF networks: (a)--(b) Forward (fw) and backward (bw) synchronization profiles $r$ vs $\lambda$ from direct simulation of a SF network of size $N=1000$ for $\epsilon=0$ and $20$. (c)--(d) Synchronization profiles $r$ vs $\lambda$ given by solution curves of Eqs.~(\ref{eq:Self01}) and (\ref{eq:Self02}) for $\epsilon=0$ and $20$. Other parameters are $\gamma=3.5$ and $k_0=10$.} \label{fig:HystSF}
\end{figure}

In Figs.~\ref{fig:HystSF}(a) and \ref{fig:HystSF}(b) we plot the forward and backward synchronization profiles from direct simulation of Eq.~(\ref{eq:Kuramoto}) for $\epsilon=0$ and $15$ on a network of size $N=1000$ constructed using the configuration model~\cite{Bekessy1972}. We note that, similar to the C. elegans network and the SE network, in the absence of disorder the transition from incoherence to synchronization is second-order, but with enough disorder the transition becomes explosive with a clear bistable regime emerging. We complement these by plotting in Fig.~\ref{fig:HystSF}(c) and \ref{fig:HystSF}(d) the solution curves of the self-consistency condition in Eqs.~(\ref{eq:Self01}) and (\ref{eq:Self02}) for the same $\epsilon$ values, calculated numerically. Again we deduce the stability of each branch using the topology of the curves, plotting stable and unstable branches in solid blue and dashed red, respectively, and results are plotted with the same horizontal axis range and stacked vertically with corresponding $\epsilon$ values of the simulation results in Figs.~\ref{fig:HystSF}(a) and \ref{fig:HystSF}(b) for easy comparison. We note first that the self-consistency condition correctly predicts a second-order transition for no disorder ($\epsilon=0$) and a first-order transition for $\epsilon=15$ via the formation of a bistable region. Again, they provide a reasonable prediction for the critical coupling values of the transitions.

\begin{figure}[t]
\centering
\epsfig{file =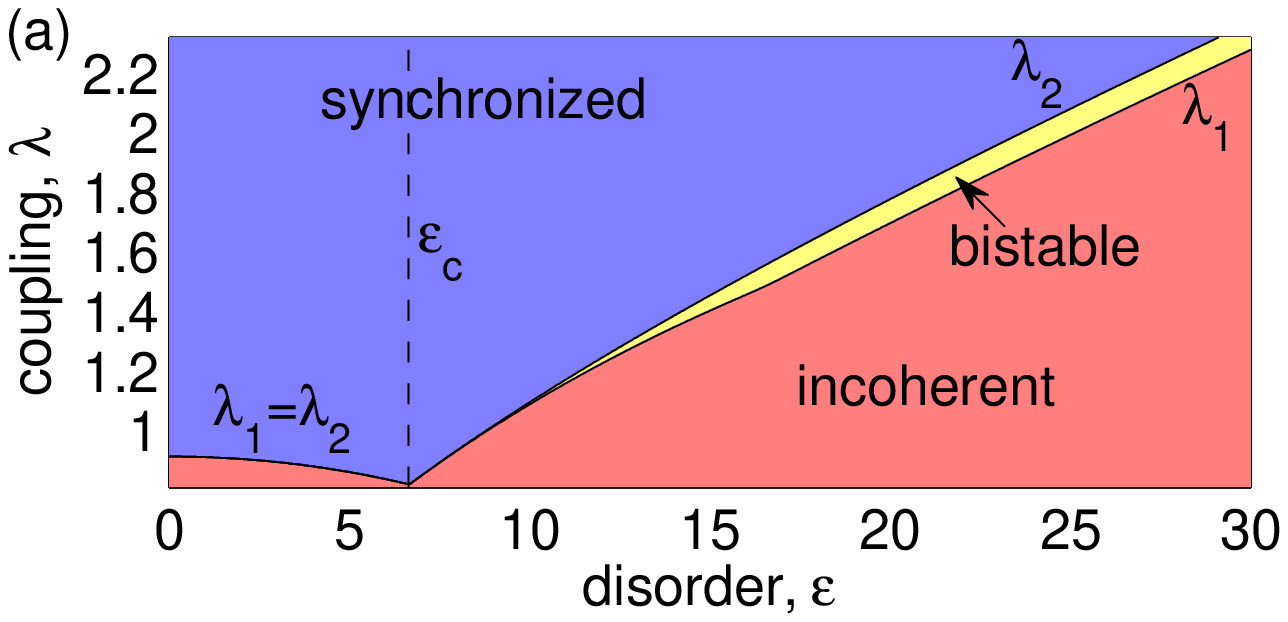, clip =,width=0.96\linewidth } \\
\epsfig{file =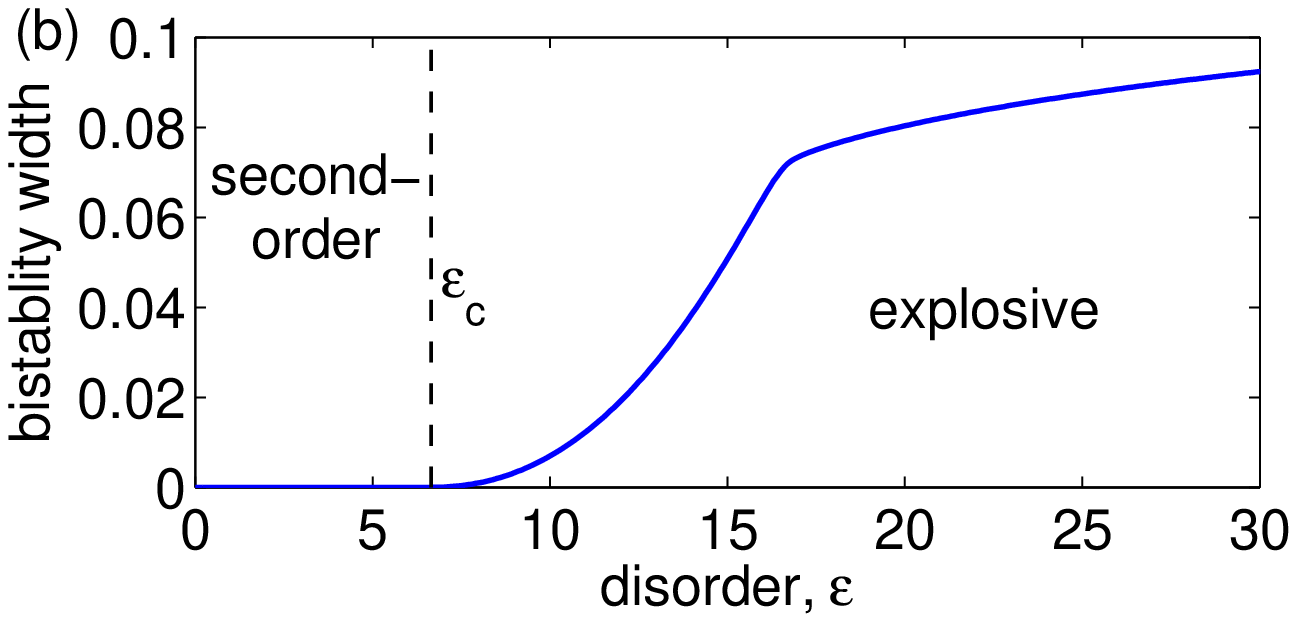, clip =,width=0.96\linewidth }
\caption{(Color online) Phase space of SF networks: (a) Phase space with critical coupling strengths $\lambda_1$ and $\lambda_2$ separating incoherent, synchronized, and bistable regions and (b) the bistability width $\lambda_2-\lambda_1$ versus $\epsilon$ computed from Eqs.~(\ref{eq:Self01}) and (\ref{eq:Self02}). Other parameters are $\gamma=3.5$ and $k_0=10$.} \label{fig:PhaseSpaceSF}
\end{figure}

Finally, we use Eqs.~(\ref{eq:Self01}) and (\ref{eq:Self02}) to compute the phase space and bistability width of SF networks. In Fig.~\ref{fig:PhaseSpaceSF}(a) we plot the phase space for SF networks with $\gamma=3.5$ and $k_0=10$, and in Fig.~\ref{fig:PhaseSpaceSF}(b) we plot the bistability width $\lambda_2-\lambda_1$ as a function of $\epsilon$. We find that the results are qualitatively similar to those of SE networks. Namely, for small disorder values $\epsilon$ the transition from incoherent to synchronized regions is second-order with $\lambda_1=\lambda_2$ until at $\epsilon_c\approx6.65$ (denoted with a vertical dashed line), a bistable region is born with $\lambda_1<\lambda_2$.

\section{Discussion}\label{sec5}

In this paper we have studied explosive synchronization in networks with degree-frequency correlations. Many recent studies have investigated explosive synchronization, however in each case strong conditions are imposed on the network to engineer first-order phase transitions, for instance strong structural heterogeneity. Here we have focused on mildly heterogeneous networks and shown that, with the addition of quenched disorder to the oscillators' frequencies, explosive synchronization can be induced in networks that do not display explosive synchronization in the absence of quenched disorder. We have used numerical and analytical tools to study this phenomenon on the real C. elegans neural network as well as two classes of synthetic networks: stretched exponential and scale-free networks. 

With the addition of quenched disorder, the oscillators' frequencies in the resulting network are not precisely determined by, but rather correlated with, local structural properties. This is a property one might expect to find in many real-world networks. Importantly, our results show that the phenomenon of explosive synchronization is not only robust to, but additionally promoted by, this natural mechanism. We emphasize here that the quenched disorder we consider is distinct from temporal fluctuations, and thus these results should be viewed as complimentary to, rather than an extension of, other studies of stochastic resonance~\cite{Ullner2003PRL,Perc2007PRE,Sagues2007RMP} and noise-induced phenomena~\cite{Lai2013PRE}.

\acknowledgements
This work has been partially supported by the Spanish DGICYT grant FIS2012-38266, FwET project MULTIPLEX (317532), and the James S. McDonnell Foundation. A.A. acknowledges the ICREA Academia.

\begin{appendix}

\section{C. elegans neural network dataset}\label{appA}

\begin{figure}[t]
\centering
\epsfig{file =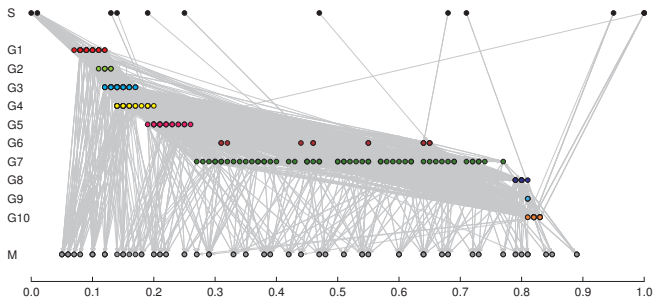, clip =,width=\linewidth }
\caption{C. elegans layout of sensorial (S), motor (M) and non-pharyngeal neuronal cells, and its connectivity. Neurons in the same ganglia are given the same vertical offset for clarity: G1) anterior ganglion, G2) dorsal ganglion, G3) lateral ganglion, G4) ventral ganglion, G5) retrovesicular ganglion, G6) posterolateral ganglion, G7) ventral cord neuron group, G8) pre-anal ganglion, G9) dorsorectal ganglion, G10) lumbar ganglion. The bottom ruler shows the longitudinal assigned coordinates, with values 0.0 and 1.0 for the head and tail of the worm respectively.}
\label{fig:CEnet}
\end{figure}

The nematode {\em Caenorhabditis Elegans} is an example of an organism where experimental research has allowed us to gain insight and understand the mechanisms underlying a whole animal's behavior at both the molecular and cellular levels~\cite{Wood1988,Schafer2005CB}. It has been extensively studied to understand particular biological phenomena, with the expectation that discoveries made in this organism will provide insight into the workings of other organisms, and thus can be considered a model organism. 


The structural anatomy of C. elegans is basically that of a cylinder around 1 millimeter in length and 0.1 millimeter in diameter. In the following, we will use the common hypothesis of study of this animal as a one dimensional entity. We are interested in its neuronal system, in particular the position along the body of the different neurons and their interconnections. The current work uses the public data found in~\cite{Wormatlas}. The construction of this data set started with the work by Albertson et al., and White et al.~\cite{Albertson1976,White1986}, and has been contributed by many authors since then, in the multimedia project {\em Wormatlas}~\cite{Wormatlas}. The particular wiring diagram we use was revised and completed by Chen et al.~\cite{Chen2006PNAS} using other valuable sources~\cite{Durbin1987,Achacoso1992}. The wiring information we have used is structured in four parts: connectivity data between neurons, neuron description, neuron connections to sensory organs and body muscles, and neuronal lineage. The architecture of the nervous system of C. elegans shows a bilaterally symmetric body plan. With a few exceptions, neurons in  C. elegans have a simple uni- or bipolar morphology that is typical for invertebrates. Synapses between neurons are usually formed {\em en passant} and each cell has multiple presynaptic regions dispersed along the length of the axon.

The neuronal network connectivity of the C. elegans can be represented as a weighted adjacency matrix of 279 nonpharyngeal neurons, out of a total of 302 neurons (pharyngeal neurons are not considered in this work because they are not reported in the above mentioned database). The abstraction at this point consists in to assume that the nervous system of the C. elegans can be modeled as a network, where nodes represent the center of the cell bodies, and the links represent synapses, see Fig.~\ref{fig:CEnet}. The order and nomenclature of the neurons in the matrix follows that of~\cite{Achacoso1992}, for a detailed biological record of the dataset see~\cite{Wormatlas}. The position of neurons has been defined in the data set as follows: i) neuron location is considered at the center of the cell body projected onto the anterior--posterior axis of the worm, ii) a neuron is assumed to make a single connection to a given sensory organ, iii) the position of each muscle is defined as the midpoint between anterior and posterior extremities of the sarcomere region, and iv) there is a lack of data specifying the location of individual synapses in the worm.

\section{Contribution of drifting oscillators}\label{appB}

Here we extend the theory outlined in Sec.~\ref{sec3} and specifically consider the contribution of the drifting oscillators to the local order parameters. Recall that in neglecting the drifting oscillators, we considered only those that satisfied the phase-locking criteria $|\omega_i-\Omega|\le\lambda r_i$. Here we consider also oscillators that satisfy $|\omega_i-\Omega|>\lambda r_i$ so that no fixed point exists for Eq.~(\ref{eq:Theory03}). It is important to note that the contribution of the drifting population can be in principle non-zero since each oscillator $\phi_i$ spends more (less) time near the minimum (maximum) of $|\dot{\phi}_i|$. We begin by introducing the density function $\rho(\phi;\omega_i,r_i)$ representing the probability of finding $\phi_i$ at $\phi$ is given by
\begin{align}
\rho(\phi,\omega_i,r_i) = \frac{\sqrt{(\omega_i-\Omega)^2-\lambda^2r_i^2}}{2\pi(\omega_i-\Omega-\lambda r_i\sin\phi)}.\label{eq:B01}
\end{align}
Here we have assumed $\psi_i=0$, which can be done without loss of generality by imposing a suitable rotation in initial conditions.

Following our analysis in the main text and taking into account the main approximation, most notably that (i) each local order parameter is proportional to its degree, i.e., $r_i=rk_i$, and that (ii) each average phase is approximately equal, i.e., $\psi_i=\psi_j$ for all $(i,j)$ pairs, we arrive at 
\begin{align}
rk_i&=\sum_{|\omega_j-\Omega|\le\lambda rk_j}A_{ij}\left[\sqrt{1-\left(\frac{\omega_j-\Omega}{\lambda rk_j}\right)^2}+i\frac{\omega_j-\Omega}{\lambda rk_j}\right]\nonumber\\&+\sum_{|\omega_j-\Omega|>\lambda rk_j} A_{ij}\int_0^{2\pi}e^{i\phi}\rho(\phi;\omega_j,rk_j)d\phi,\label{eq:B02}
\end{align}
which is the same as Eq.~(\ref{eq:Theory05}) of the main text plus the addition of the final term describing the drifting contributions. Summing over $i$, dividing by $N$ and rearranging yields the following complicated set of equations.
\begin{widetext}
\begin{align}
r &= \frac{\langle k\rangle^{-1}}{N}\sum_{|\omega_j-\Omega|\le\lambda rk_j}k_j\sqrt{1-\left(\frac{\omega_j-\Omega}{\lambda rk_j}\right)^2}+\frac{\langle k\rangle^{-1}}{N}\sum_{|\omega_j-\Omega|>\lambda rk_j}k_j\int_0^{2\pi}\cos\phi\rho(\phi;\omega_j,rk_j)d\phi,\label{eq:B3}\\
\Omega &= \frac{\sum_{|\omega_j-\Omega|\le\lambda rk_j}\omega_j+\lambda r\sum_{|\omega_j-\Omega|>\lambda rk_j}k_j\int_0^{2\pi}\sin\phi\rho(\phi;\omega_j,rk_j)d\phi}{\sum_{|\omega_j-\Omega|\le\lambda rk_j}1}.\label{eq:B4}
\end{align}
\end{widetext}
Equations~(\ref{eq:B3}) and (\ref{eq:B4}) are the analogous forms of Eqs.~(\ref{eq:Theory06}) and (\ref{eq:Theory07}) in the text, and again represent a simple extension accounting for the drift oscillators. We note, however, that due to the dependence of $\rho$ on $\Omega$, Eq.~(\ref{eq:B4}) defines $\Omega$ implicitly. Finally, the integral versions of Eqs.~(\ref{eq:B3}) and (\ref{eq:B4}) are given by
\begin{widetext}
\begin{align}
r &= \langle k\rangle^{-1}\iint_{|\omega-\Omega|\le\lambda rk}P(k,\omega)k\sqrt{1-\left(\frac{\omega-\Omega}{\lambda rk}\right)^2}d\omega dk+\langle k\rangle^{-1}\iint_{|\omega-\Omega|>\lambda rk}P(k,\omega)k\left[\int_0^{2\pi}\cos\phi\rho(\phi;\omega,rk)d\phi\right]d\omega dk,\label{eq:B5}\\
\Omega &= \frac{\iint_{|\omega-\Omega|\le\lambda rk}P(k,\omega)\omega d\omega dk+\lambda r\iint_{|\omega-\Omega|>\lambda rk}P(k,\omega)k\left[\int_0^{2\pi}\sin\phi\rho(\phi;\omega,rk)d\phi\right]d\omega dk}{\iint_{|\omega-\Omega|\le\lambda rk}P(k,\omega)d\omega dk}.\label{eq:B6}
\end{align}
\end{widetext}
Equations~(\ref{eq:B5}) and (\ref{eq:B6}) are those analogous to Eqs.~(\ref{eq:Self01}) and (\ref{eq:Self02}) in the main text, taking into account the contribution of drifting oscillators. We note, however, that numerical investigation suggest that the drifting contributions accounted for in Eqs.~(\ref{eq:B5}) and (\ref{eq:B6}) are second-order in comparison to the locked contribution. Thus, for the sake of simplicity all theoretical curves presented in this paper are calculated using Eqs.~(\ref{eq:Self01}) and (\ref{eq:Self02}) in the main text.

\section{Numerical investigations}\label{appC}

We now present some numerical investigations supporting some approximations and remarks made in Section~\ref{sec3}. First, we investigate the main approximations made in deriving the self-consistency condition. Second, we compare the integral and sum versions of the self-consistency condition.

\subsection*{Validation of approximations}

\begin{figure}[b]
\centering
\epsfig{file =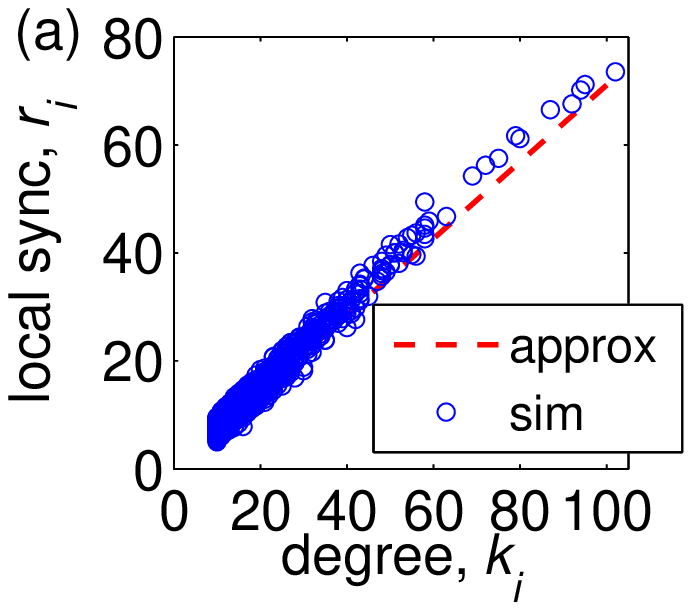, clip =,width=0.48\linewidth }
\epsfig{file =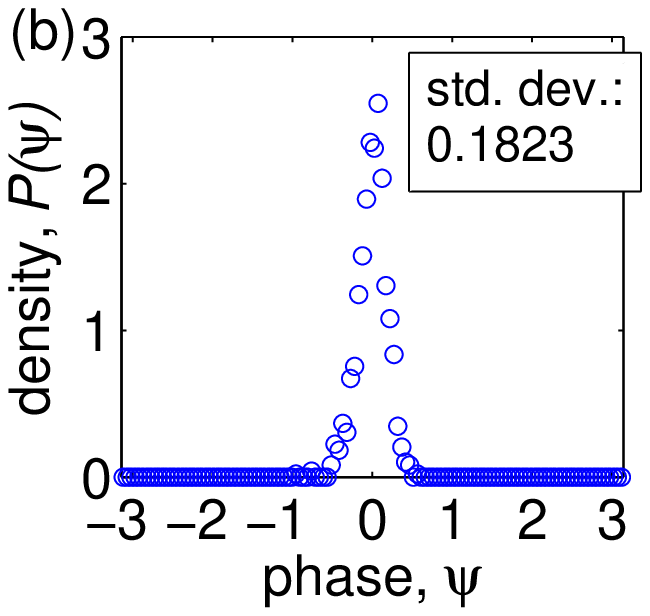, clip =,width=0.48\linewidth }
\caption{(Color online) Numerical verification of approximations: (a) Local order parameters $r_i$ vs $k_i$ compared to the approximation $rk_i$. (b) Distribution of average phases $\psi_i$. The network is SE of size $N=1000$ with $\mu=5$, $\beta=0.7$, and $k_0=10$. Other parameters are $\lambda=1.5$ and $\epsilon=15$.} \label{fig:Approx}
\end{figure}

Here we present some numerical experiments supporting the validity of approximations made int he theoretical derivation of Eqs.~(\ref{eq:Self01}) and (\ref{eq:Self02}) as well ass Eqs.~(\ref{eq:B5}) and (\ref{eq:B6}). In particular, we address the following two major approximations regarding the local order parameters:
\begin{enumerate}[(i)]
\item that each local order parameter is approximately proportional to its corresponding nodal degree, i.e., $r_i\approx r k_i$ given $r\in[0,1]$, and
\item that each average phase $\psi_i$ is approximately equal, i.e., $\psi_i\approx\psi_j$ for all $(i,j)$ pairs.
\end{enumerate}
We validate these approximations by performing direct simulations for a SE network of size $N=1000$ with $\mu=5$, $\beta=0.7$, and $k_0=10$ (the same parameter choices used in the main text), and setting $\lambda=1.5$ and $\epsilon=15$ and extracting each $r_i$ and $\psi_i$. In Fig.~\ref{fig:Approx} (a) we plot each local order parameter $r_i$ vs $k_i$ (blue circles) compared to the approximation $rk_i$, where $r$ is the global Kuramoto order parameter calculated from the simulation (in this simulation we found $r=0.697$). We note an excellent agreement between the local order parameters extracted from the simulation and the approximation. Next, we calculate the probability distribution of average phases $\psi$ (shifted to set the mean phase to zero) and plot the results in panel (b). We note that the phases are all tightly packed near the mean of zero, and calculate the standard deviation in the phases to be $0.1823$.

\subsection*{Integrals vs. sums}

\begin{figure}[t]
\centering
\epsfig{file =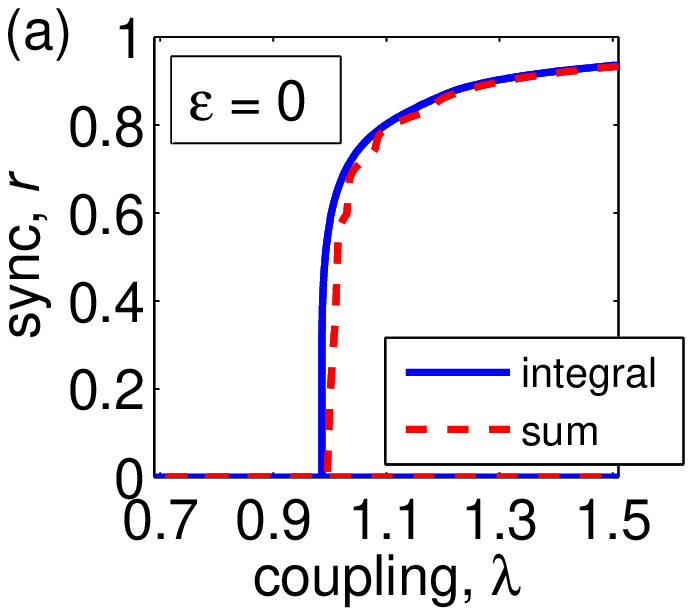, clip =,width=0.48\linewidth }
\epsfig{file =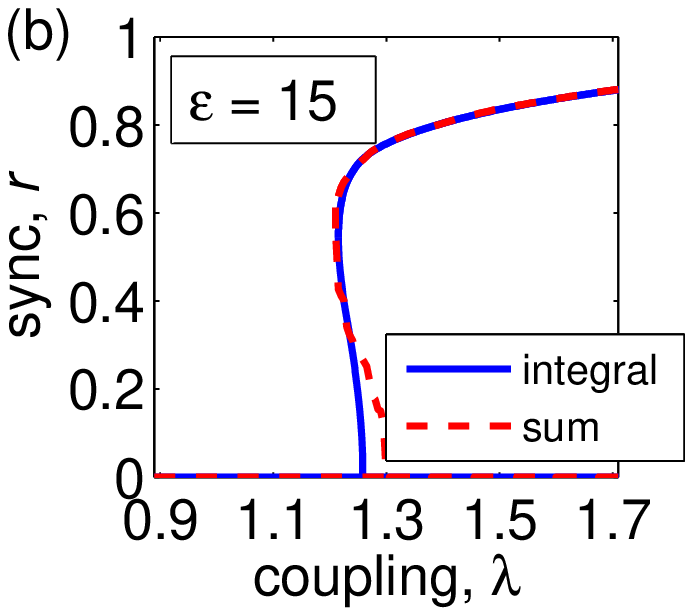, clip =,width=0.48\linewidth }
\caption{(Color online) Integrals vs. sums: (a)--(b) Comparison of the solution of the integral (solid blue) and sum (dashed red) self consistency conditions [Eqs.~(\ref{eq:Self01}) and (\ref{eq:Self02}) vs. Eqs.~(\ref{eq:Theory06}) and (\ref{eq:Theory07})] for SE networks with $\epsilon=0$ and $15$. The summation solutions are calculated from a degree-frequency sequence $\{(k_i,\omega_i)\}_{i=1}^N$ for a network of size $N=1000$. Other parameters are $\mu=5$, $\beta=0.7$, and $k_0=10$.} \label{fig:Compare}
\end{figure}

Next we compare the results from using the integral and sum versions of the self-consistency condition, i.e., solving Eqs.~(\ref{eq:Self01}) and (\ref{eq:Self02}) vs. solving Eqs.~(\ref{eq:Theory06}) and (\ref{eq:Theory07}). We consider SE networks and compare the solutions of Eqs.~(\ref{eq:Self01}) and (\ref{eq:Self02}), which we already computed and used in Figs.~\ref{fig:HystSE}(c) and \ref{fig:HystSE}(d), to solutions of Eqs.~(\ref{eq:Theory06}) and (\ref{eq:Theory07}) computed by extracting the degree-frequency sequence $\{(k_i,\omega_i)\}_{i=1}^N$ from the simulation presented in Fig.~\ref{fig:HystSE}(a) and \ref{fig:HystSE}(b). Parameters are $N=1000$, $\mu=5$, $\beta=0.7$, $k_0=10$, and $\epsilon=0$ and $15$. In Fig.~\ref{fig:Compare}(a) and \ref{fig:Compare}(b) we plot the resulting curves from solving the integral equations (solid blue) to the summation equations (dashed red) for $\epsilon=0$ and $15$, respectively. We note that the solution curves match up closely for both $\epsilon=0$ and $\epsilon=15$. These results highlight the utility of the integrals [Eqs.~(\ref{eq:Self01}) and (\ref{eq:Self02})] since they require only the joint distribution, as opposed to the sums [Eqs.~(\ref{eq:Theory06}) and (\ref{eq:Theory07})] which require the whole sequence of degree-frequency pairs.

\end{appendix}

\bibliographystyle{plain}

\begin{thebibliography}{99}
\bibitem{Dorogovtsev2008RMP} S. N. Dorogovtsev, A. V. Goltsev, and J. F. F. Mendes, J. F. F., Rev. Mod. Phys. {\bf 80}, 1275 (2008).
\bibitem{Achlioptas2009Science} D. Achlioptas, R. D'Souza, and J. Spencer, Science {\bf 323}, 1453 (2009).
\bibitem{daCosta2010PRL} R. A. da Costa, S. N. Dorogovtsev, A. V. Goltsev, and J. F. F. Mendes, Phys. Rev. Lett. {\bf 105}, 255701 (2010).
\bibitem{Riordan2011Science} O. Riordan and L. Warnke, Science {\bf 333}, 322 (2011).
\bibitem{Arenas2008PR} A. Arenas, A. D\'{i}az-Guilera, J. Kurths, Y. Moreno, and C. Zhou, Phys. Rep. {\bf 469}, 93 (2008).
\bibitem{Strogatz2003} S. H. Strogatz, {\it Sync: the Emerging Science of Spontaneous Order} (Hypernion, 2003).
\bibitem{Pikovsky2003} A. Pikovsky, M. Rosenblum, and J. Kurths, {\it Synchronization: A Universal Concept in Nonlinear Sciences} (Cambridge University Press, 2003).
\bibitem{Buck1988QRB} J. Buck, Q. Rev. Biol. {\bf 63}, 265 (1988).
\bibitem{Glass1988} L. Glass and M. C. Mackey, {\it From Clocks to Chaos: The Rhythms of Life} (Princeton University Press, Princeton, 1988).
\bibitem{Motter2013NaturePhys} A. E. Motter, S. A. Myers, M. Anghel, and T. Nishikawa, Nature Phys. {\bf 9}, 191 (2013).
\bibitem{Strogatz2005Nature} S. H. Strogatz, D. M. Abrams, A. McRobie, B. Eckhardt, and E. Ott, Nature (London) {\bf 438}, 43 (2005).
\bibitem{Prindle2012Nature} A. Prindle, P. Samayoa, I. Razinkov, T. Danino, L. S. Tsimring, and J. Hasty, Nature {\bf 481}, 39 (2012).
\bibitem{Kuramoto1984} Y. Kuramoto, {\it Chemical Oscillations, Waves, and Turbulence} (Springer, New York, 1984).
\bibitem{GomezGardenes2011PRL} J. G\'{o}mez-Garde\~{n}es, S. G\'{o}mez, A. Arenas, and Y. Moreno, Phys. Rev. Lett. {\bf 106} 128701 (2011).
\bibitem{Barabasi1999Science} A.-L. Barab\'{a}si and R. Albert, Science {\bf 286}, 509 (1999).
\bibitem{Pazo2005PRE} D. Paz\'{o}, Phys. Rev. E {\bf 72}, 046211 (2005).
\bibitem{Skardal2013EPL} P. S. Skardal, J. Sun, D. Taylor, and J. G. Restrepo, Europhys. Lett. {\bf 101}, 20001 (2013).
\bibitem{Liu2013EPL} W. Liu, Y. Wu, J. Xiao, and M. Zhan, Europhys. Lett. {\bf 101}, 38002 (2013).
\bibitem{Sonnenschein2013EPJB} B. Sonnenschein, F. Sagu\'{e}s, and L. Schimansky-Geier, Eur. Phys. J. B {\bf 86}, 12 (2013).
\bibitem{Coutinho2013PRE} B. C. Coutinho, A. V. Goltsev, S. N. Dorogovtsev, and J. F. F. Mendes, Phys. Rev. E {\bf 87}, 032106 (2013).
\bibitem{Leyva2012PRL} I. Leyva, R. Sevilla-Escoboza, J. M. Buld\'{u}, I. Sendi\~{n}a-Nadal, J. G\'{o}mez-Garde\~{n}es, A. Arenas, Y. Moreno, S. G\'{o}mez, R. Jaimes-Re\'{a}tegui, and S. Boccaletti, Phys. Rev. Lett. {\bf 108}, 168702 (2012).
\bibitem{Peron2012PRE1} Thomas Kaue Dal Maso Peron and F. A. Rodrigues, Phys. Rev. E {\bf 86}, 016102 (2012).
\bibitem{Peron2012PRE2} Thomas Kaue Dal Maso Peron and F. A. Rodrigues, Phys. Rev. E {\bf 86}, 056108 (2012).
\bibitem{Chen2013Chaos} H. Chen, G. He, F. Huang, C. Shen, and Z. Hou, Chaos {\bf 23}, 033124 (2013).
\bibitem{Ji2013PRL} P. Ji, T. K. DM. Peron, P. J. Menck, F. A. Rodrigues, and J. Kurths, Phys. Rev. Lett. {\bf 110}, 218701 (2013).
\bibitem{Leyva2013SR} I. Leyva, A. Navas, I. Sendi\~{n}a-Nadal, J. A. Almendral, J. M. Buld\'{u}, M. Zanin, D. Papo, and S. Boccaletti, Sci. Rep. {\bf 3}, 1281 (2013).
\bibitem{Leyva2013PRE} I. Leyva, I. Sendi\~{n}a-Nadal, J. A. Almendral, A. Navas, S. Olmi, and S. Boccaletti, Phys. Rev. E {\bf 88}, 042808 (2013).
\bibitem{Li2013PRE} P. Li, K. Zhang, X. Xu, J. Zhang, and M. Small, Phys. Rev. E {\bf 87}, 042803 (2013).
\bibitem{Su2013EPL} G. Su, Z. Ruan, S. Guan, and Z. Liu, Europhys. Lett. {\bf 103}, 48004 (2013).
\bibitem{Zhang2013PRE} X. Zhang, X. Hu, J. Kurths, and Z. Liu, Phys. Rev. E {\bf 88}, 010802(R) (2013).
\bibitem{Zhu2013PRE} L. Zhu, L. Tian, and D. Shi, Phys. Rev. E {\bf 88}, 042921 (2013).
\bibitem{Zou2014PRL} Y. Zou, T. Pereira, M. Small, Z. Liu, and J. Kurths, Phys. Rev. Lett. {\bf 112}, 114102 (2014).
\bibitem{Chen2006PNAS} B. L. Chen, D. H. Hall, and D. B. Chklovskii, Proc. Natl. Acad. Sci. U.S.A. {\bf 103}, 4723 (2006).
\bibitem{Ichinomiya2004PRE} T. Ichinomiya, Phys. Rev. E {\bf 70}, 026116 (2004).
\bibitem{Restrepo2005PRE} J. G. Restrepo, E. Ott, and B. R. Hunt, Phys. Rev. E {\bf 71}, 036151 (2005).
\bibitem{Clauset2009SIAM} A. Clauset, C. R. Shalizi, and M. E. J. Newman, SIAM Rev. {\bf 51}, 661 (2009).
\bibitem{Erdos1960} P. Erd\H{o}s and A. R\'{e}nyi, Pub. Math. Inst. Hung. Acad. Sci. {\bf 5}, 17 (1960).
\bibitem{Bekessy1972} A. Bekessy, P. Bekessy, and J. Komlos, Stud. Sci. Math. Hung. {\bf 7}, 343 (1972).
\bibitem{Ullner2003PRL} E. Ullner, A. Zaikin, J. Garc\'{i}a-Ojalvo, and J. Kurths, Phys. Rev. Lett. {\bf 91},180601 (2003).
\bibitem{Perc2007PRE} M. Perc, Phys. Rev. E {\bf 76}, 066203 (2007).
\bibitem{Sagues2007RMP} F. Sagu\'{e}s, J. M. Sancho, and J. Garc\'{i}a-Ojalvo, Rev. Mod. Phys. {\bf 79}, 829 (2007).
\bibitem{Lai2013PRE} Y. M. Lai and M. A. Porter, Phys. Rev. E {\bf 88}, 012905 (2013).
\bibitem{Wood1988} W. B. Wood, {\it The nematode Caenorhabditis elegans} (Cold Spring Harbor Laboratory: Cold Spring Harbor, 1988)
\bibitem{Schafer2005CB} W. R. Schafer, Curr. Biol. {\bf 15}, R723 (2005).
\bibitem{Wormatlas} http://www.wormatlas.org/neuronalwiring.html
\bibitem{Albertson1976} D. G. Albertson and J. N. Thomson, Phil. Trans. R. Soc. London {\bf 275}, 299 (1976).
\bibitem{White1986} J. G. White, E. Southgate, J. N. Thompson, and S. Brenner, Phil. Trans. Royal Soc. London {\bf 314}, 1 (1986).
\bibitem{Durbin1987} R. M. Durbin, Ph.D. thesis, University of Cambridge, 1987.
\bibitem{Achacoso1992} T. B. Achacoso and W. S. Yamamoto, {\it AY's Neuroanatomy of C. elegans for Computation} (CRC Press, 1992).

\end{thebibliography}

\end{document}